\documentclass[conference]{IEEEtran}

\ifCLASSINFOpdf
\else
\fi
\usepackage{graphicx}

\usepackage{amsmath}
\usepackage{siunitx}
\usepackage{cleveref}
\begin{document}
%
\title{Performance of  a Highly Granular Scintillator-SiPM Based  Hadron Calorimeter Prototype  in Strong Magnetic Fields}

\author{\IEEEauthorblockN{Christian Graf}
\IEEEauthorblockA{Max-Planck-Institute for Physics, Munich, Germany}
\IEEEauthorblockA{Email: cgraf@mpp.mpg.de}

\IEEEauthorblockA{for the CALICE Collaboration}}


%


\maketitle

\begin{abstract}
Within the CALICE collaboration, several concepts for the hadronic calorimeter
of a future linear collider detector are studied. After having demonstrated the
capabilities of the measurement methods in "physics prototypes", the focus now
lies on improving their implementation in "engineering prototypes", that are
scalable to the full linear collider detector. The Analog Hadron Calorimeter
(AHCAL) concept is a sampling calorimeter of tungsten or steel absorber plates
and plastic scintillator tiles read out by silicon photomultipliers (SiPMs) as
active material. The front-end chips are integrated into the active layers of
the calorimeter and are designed for minimizing power consumption by rapidly
cycling the power according to the beam structure of a linear accelerator. The
versatile electronics allows the prototype to be equipped with different types
of scintillator tiles and SiPMs. A prototype with  $\sim$2200 channels, equipped
with several types of scintillator tiles and SiPMs, was tested with muons
and electrons in a 1.5 T magnet at the CERN SPS in May 2017 to establish the
operational stability with power pulsing and the overall detector performance
in a magnetic field.

\end{abstract}


%
\IEEEpeerreviewmaketitle

\section{Introduction}
The physics at future high-energy lepton colliders, with its requirement for a
jet energy reconstruction with unprecedented precision, is one of the primary
motivations for the development of highly granular calorimeters by the CALICE
collaboration. The detector concepts for the International Linear Collider (ILC)
and the Compact Linear Collider (CLIC) rely on Particle Flow Algorithms (PFA)
\cite{Thomson:2009rp}, which are capable of achieving the required resolution.
This event reconstruction technique requires highly granular calorimeters to
deliver optimal performance.

One of the technologies developed within CALICE is the Analog Hadron Calorimeter
AHCAL \cite{Adloff:2010hb}. It is based on active elements consisting of
\SI{3 x 3}{\centi\metre} plastic scintillator tiles individually read out by
silicon photomultipliers (SiPMs) in a steel absorber structure with approximately
\SI{20}{\milli\metre} of absorber material between each active layer. A full
"physics prototype", which has been extensively tested in particle beams at
DESY, CERN and Fermilab, has demonstrated the capabilities of this technology,
achieving competitive single hadron energy resolution \cite{Adloff:2012gv} and
the two-particle separation required for good PFA performance \cite{Adloff:2011ha}.
The prototype was also successfully tested with tungsten absorbers.

\section{The CALICE Analog Hadron Calorimeter Engineering Prototype}
\begin{figure}
\centering
\includegraphics[width=0.35\textwidth]{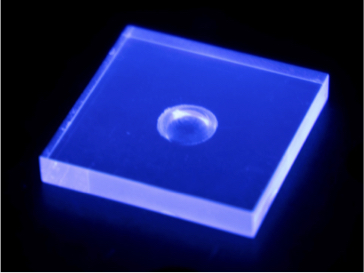}
\caption{CALICE AHCAL scintillator tile, with central dimple at the position of the photon sensor to achieve uniform response over the full area of the tile.}
\label{fig:Tile}
\end{figure}

With the establishment of the viability of the AHCAL concept, the focus is now
shifting from the study of the physical performance characteristics of such a
detector to the demonstration of the detector performance while satisfying the
spatial constraints and scalability requirements of collider experiments. This
is the goal of the "engineering prototype". This prototype uses electronics
fully embedded in the active layers to minimize space requirements for interfaces
outside of the active detector volume, and is based on scintillator tile designs
well-suited for mass production and automatic assembly. In a first iteration of
studies, different scintillator tile designs and different photon sensors were
investigated \cite{Simon:2010hf}, before converging on a design using SMD-style SiPMs and
scintillator tiles with a central dimple, as shown in \cref{fig:Tile}, to allow fiberless coupling of the
scintillator to the SiPM \cite{Liu:2015cpe}.

\begin{figure}
\centering
\includegraphics[width=0.495\textwidth]{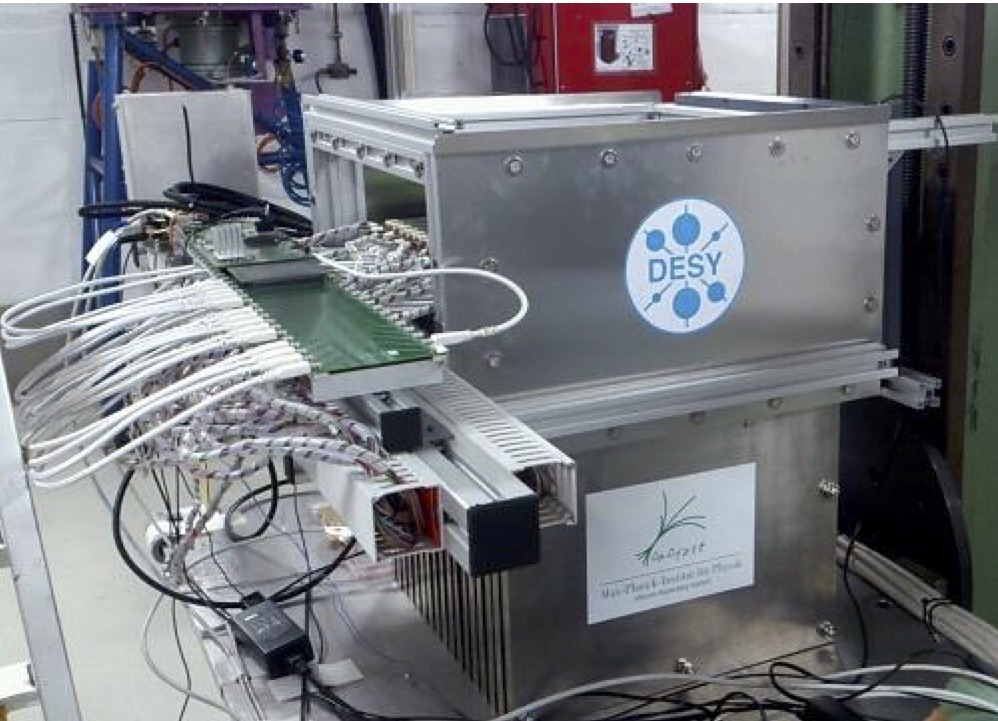}
\caption{Compact \num{15} layer AHCAL technical prototype with non-magnetic stainless steel absorber structure and \SI{36 x 36}{\centi\metre} active area.}
\label{fig:AbsorberStructure}
\end{figure}

The basic unit of the active elements is the HCAL Base Unit HBU, with a size of
\SI{36 x 36}{\centi\metre}, holding \num{144} scintillator tiles controlled by
four ASICs (SPIROC \cite{di2013spiroc}). A key element of the electronics is the capability for power-pulsed
operation to reduce the power consumption and eliminate the need for active
cooling, making use of the low duty cycle in the linear collider beam time
structure. The electronics also provide a cell-by-cell auto trigger and time
stamping on the few \si{\nano\second} level in test beam operations. In operating conditions
with shorter data-taking windows closer to the bunch train structure of linear
colliders, sub-\si{\nano\second} time resolution is available.

\begin{figure}
\centering
\includegraphics[width=0.4\textwidth]{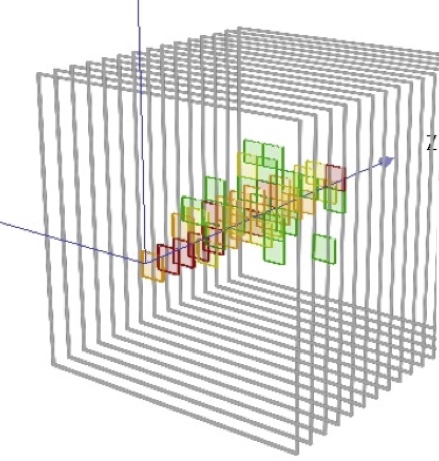}
\caption{Event display of a \SI{5}{\GeV} electron event recorded at the DESY test beam.}
\label{fig:EventDisplay}
\end{figure}

Different absorber structures are used to test the HBUs of the engineering
prototype. One of them, shown in \cref{fig:AbsorberStructure} is a compact
15 layer structure made of high-quality non-magnetic stainless steel,
housing one HBU per layer, deep enough to contain
electromagnetic showers. This allows for a precise evaluation of the detector
response with electrons. HBUs installed in this structure were recently exposed
to electron beams at DESY, illustrated by the event display shown in
\cref{fig:EventDisplay}.

\begin{figure}
\centering
\includegraphics[width=0.45\textwidth]{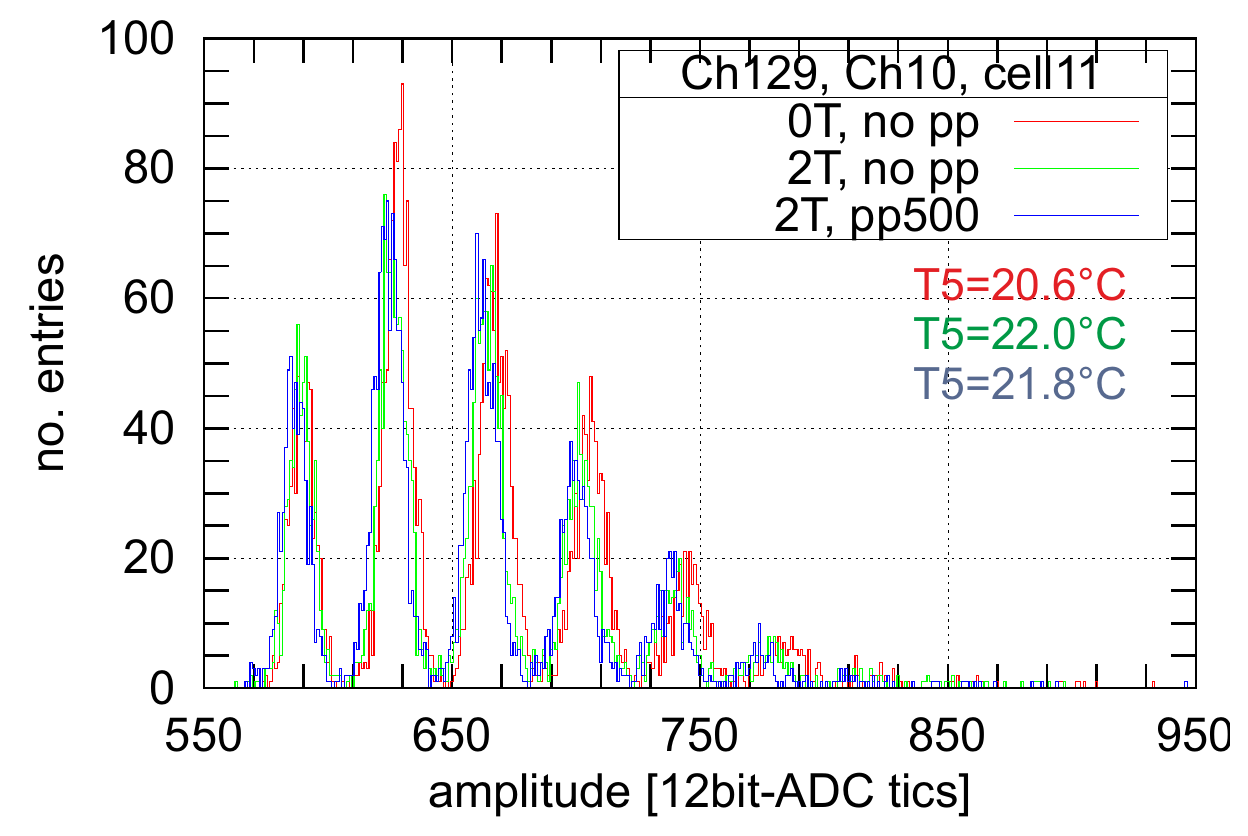}
\caption{Individual photo-electron peaks observed with low-intensity LED light without magnetic field and in a 2 T magnetic field without and with power pulsing.}
\label{fig:PPTest}
\end{figure}

To evaluate the stability of the readout with power pulsing and in the presence of
a strong magnetic field, single HBUs were tested with power pulsing enabled in a
\SI{2}{\tesla} magnetic field at DESY. \Cref{fig:PPTest} demonstrates that the photon
sensor gain remains stable with power pulsing and within a magnetic field.
The small shift in the positions of the individual photon peaks between the case
with and without magnetic field are due to the difference in temperature for the
datasets.

\section{Test-beam Campaign in May 2017}

As a next step to validate the engineering prototype a test-beam campaign was
carried out to evaluate the performance of the prototype in a magnetic field.
For this test, the compact stainless steel absorber structure, introduced above,
was used, equipped with \num{15} HBUs. The steel stack accounts for
approximately $\num{17}X_0$ and \num{1.8} nuclear interaction lengths.

In the first six layers of the prototype new HBUs equipped with
surface-mounted SiPMs from Hamamatsu are installed. The scintillators are the
injection-molded polystyrene tiles shown in \cref{fig:Tile}.
Layer \num{7} uses an HBU with previous generation surface-mounted SiPMs.
Layers \numrange{8}{15} are equipped with older HBUs showing good quality in
former test-beam campaigns using SensL SiPMs and a different tile design
optimized for the direct coupling of the photon sensor to the tile side \cite{Simon:2010hf}.

\begin{figure}
\centering
\includegraphics[width=0.43\textwidth]{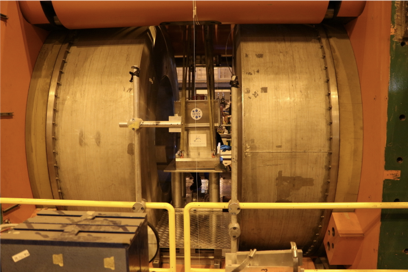}
\caption{Placement of the prototype in the magnet during the test-beam campaign.
The view is on the back of the detector with the direction of the beam facing
out of the image plane. The magnetic field is oriented from left to right.}
\label{fig:detectorInMagnet}
\end{figure}

The detector was mounted in a magnet located at the H2 beam line at
SPS (CERN), as shown in \cref{fig:detectorInMagnet}. During the ramp up
of the magnet to its design current, which provides a field at the center of the
magnet of $\SI{3}{\tesla}$, the magnet had to be turned off, because of a failure of a circuit breaker.
The detector was running without incidents up to the highest
magnetic field observed of about $\SI{2.4}{\tesla}$. Afterwards, the magnet could be operated
at a maximum magnetic field of $\SI{1.5}{\tesla}$ for the remaining duration of the test-beam campaign.
During one week of data taking, data of $\SI{120}{\GeV}$ muons and \SIlist[list-units=single]{10;20;30;40;60}{\GeV}
electrons has been recorded with and without magnetic field.

\section{Results}
The analysis of the data is still work in progress. Here, only selected first
results are presented.

\subsection{Hit Distributions}

\begin{figure}
\centering
\includegraphics[width=0.43\textwidth]{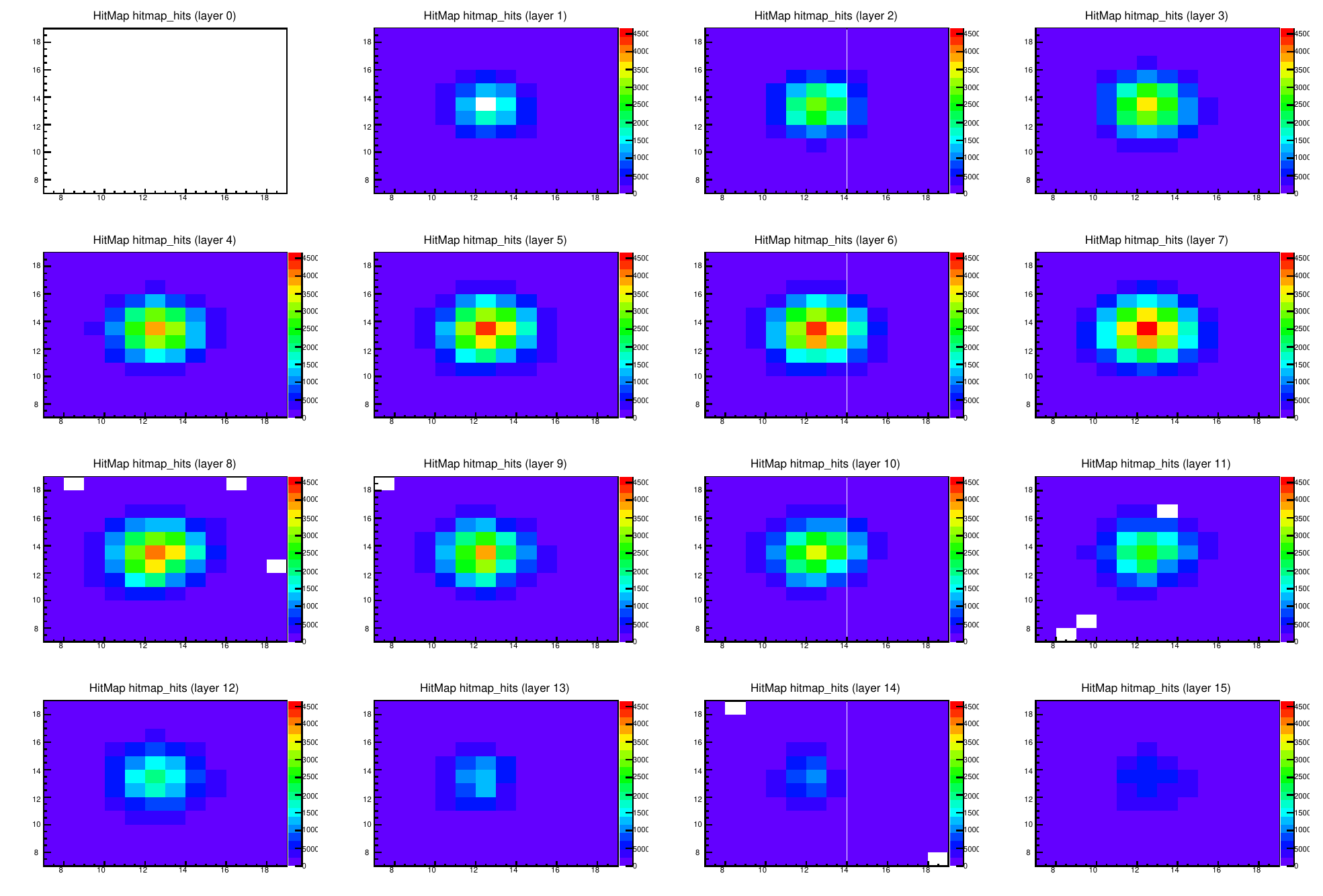}
\caption{Hit map of electromagnetic showers of \SI{10}{\GeV} electrons in all
\num{15} layers of the prototype.}
\label{fig:Hitmap_off}
\end{figure}

\begin{figure}
\centering
\includegraphics[width=0.43\textwidth]{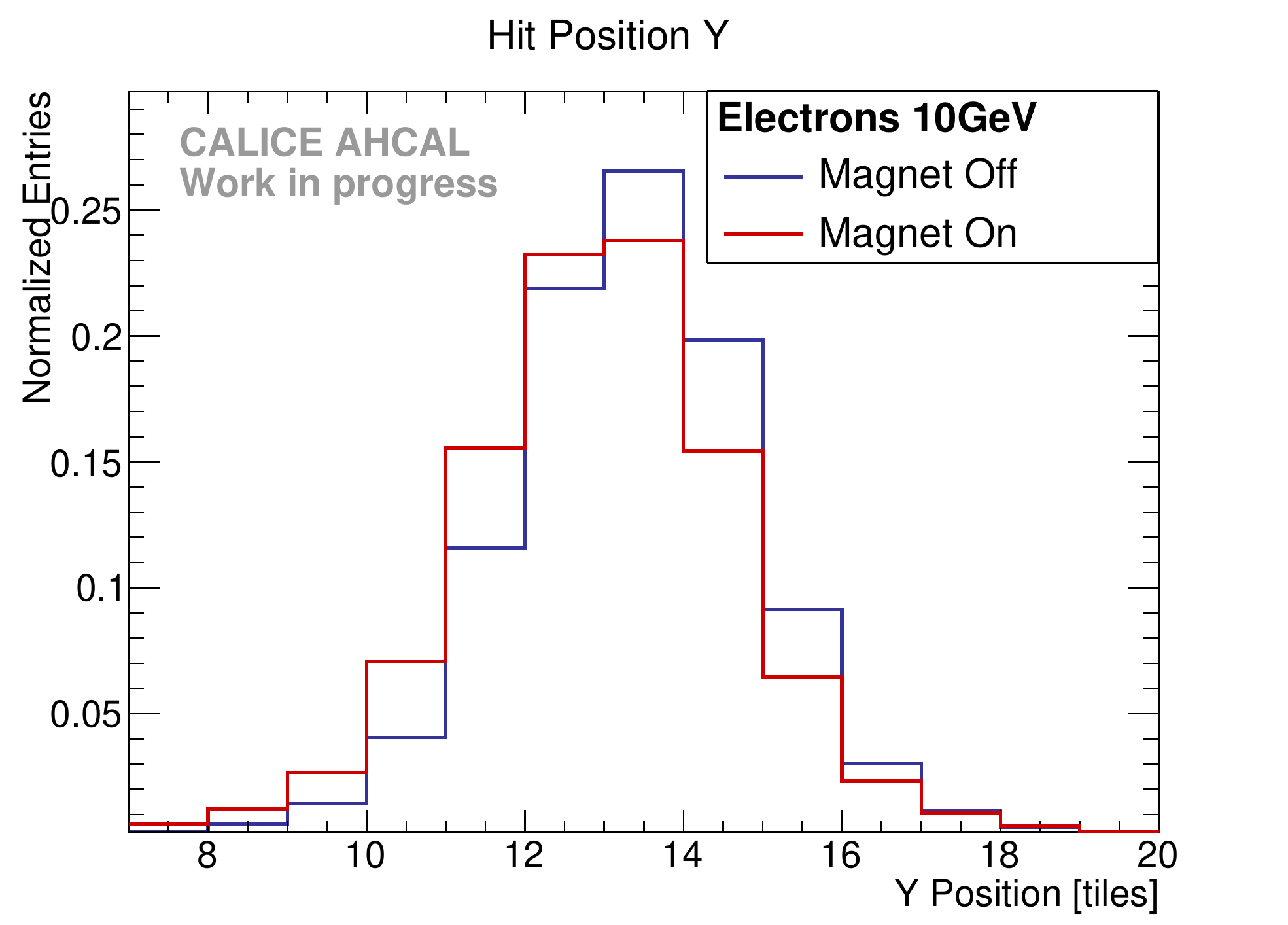}
\caption{Hit position in $y$-direction of \SI{10}{\GeV} electrons with and
without \SI{1.5}{\tesla} magnetic field.}
\label{fig:HitY}
\end{figure}

\begin{figure}
\centering
\includegraphics[width=0.43\textwidth]{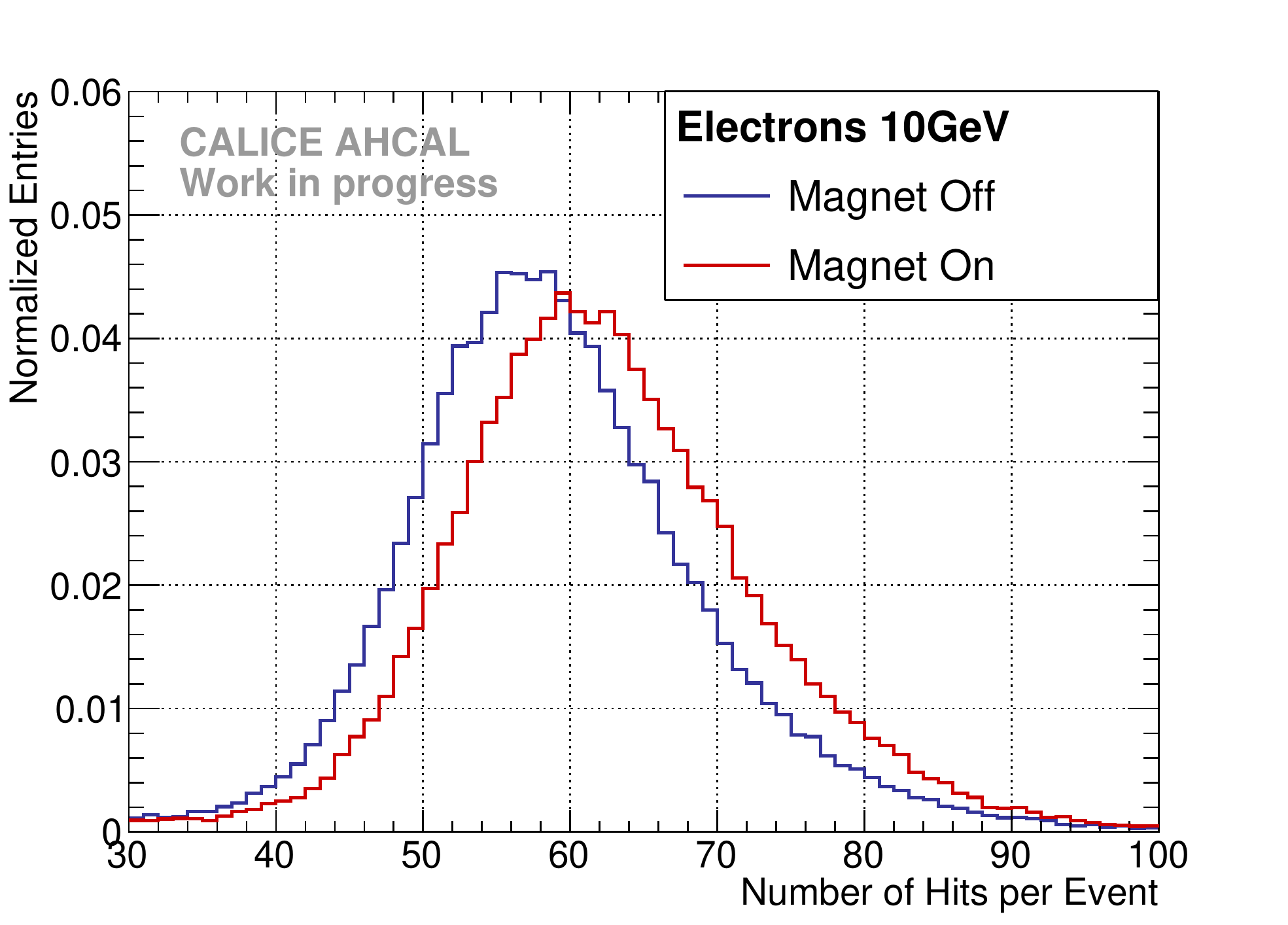}
\caption{Total number of detected hits \SI{10}{\GeV} electron events with and
without \SI{1.5}{\tesla} magnetic field.}
\label{fig:nHits}
\end{figure}

\Cref{fig:Hitmap_off} shows the distribution of hits for each layer of the
detector for \SI{10}{GeV} electrons. All readout chips are working and only very few
dead channels are observed. One can clearly see the average shape of the electromagnetic
shower with most parts confined within the detector. In general, the detector
shows a good homogeneity in hit detection efficiencies and noise rates.

\Cref{fig:detectorInMagnet} shows a view on the setup facing the back of the detector.
In this view, the magnetic field is oriented in horizontal
direction from left to right.
By comparing the distribution of the hit position with and without magnetic field
in $y$-direction, as shown in \cref{fig:HitY}, the effect of the magnetic field
on the electromagnetic shower can be observed. A shift of approximately half a tile
to smaller values on the $y$-axis is visible, as well as a slight broadening of
the shower. As expected, the hit distribution in $x$-direction shows no change
due to the magnetic field.
The broadening of the electromagnetic shower goes along with an increase in
the overall number of detected hits per event of about 10\% as shown in \cref{fig:nHits}.

\subsection{Energy Response}
It is already known from the literature that the light yield of a scintillator
is enhanced within a magnetic field \cite{bertolucci1987MagneticField}. For polystyrene
based scintillators with wavelength shifting fibers an increase in light yield
of approximately 6\% was observed in a $\SI{2}{\tesla}$ magnetic field \cite{bertoldi1997scintillators}.

For the prototype an increase of approximately 4\% is observed in the detector
response for each channel for \SI{120}{\GeV} muons in a magnetic field.
This effect adds up with the effect of more observed hits in an electromagnetic shower
and a longer geometrical path of particles bend by the magnetic field, such that the
energy response of an electromagnetic shower is increased by roughly 10\% in
a magnetic field.

\subsection{Muon Time Resolution}
\begin{figure}
\centering
\includegraphics[width=0.40\textwidth]{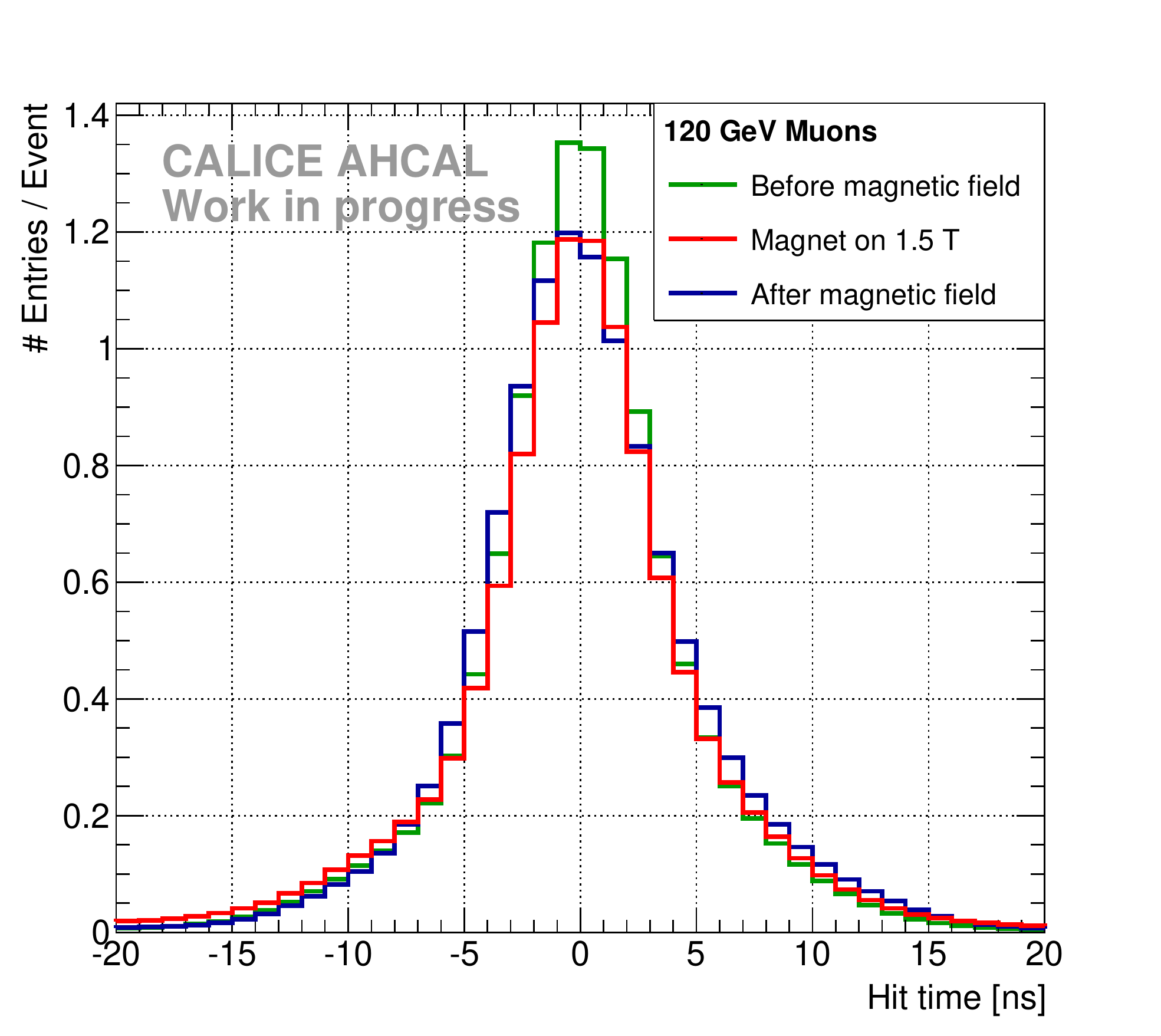}
\caption{Time distribution of \SI{120}{\GeV} muons in runs before, during and after
the magnetic field.}
\label{fig:timeDistribution}
\end{figure}
In order to check the stability of time measurements with a magnetic field the
time distribution of muons is investigated. \Cref{fig:timeDistribution}
shows the time distribution of \SI{120}{GeV} muon data of runs taken before, during and
after the magnetic field. The distribution stays stable except for a slight broadening
that is observed during and after the magnetic field. Wether this is due to an
effect of the magnetic field in the electronics, or a specific time dependent effect
within certain runs, induced for example by changing environmental conditions
needs to be investigated.

\section{Full Prototype Construction}

A full hadronic prototype of the AHCAL with approximately \num{23000} electronic
channels, organized in \num{160} HBUs on \num{40} layers, is currently under construction and
will prove the scalability of the
technology to a significant part of a full detector. It uses
Hamamatsu MPPC S13360-1325PE photon sensors and the injection-molded polystyrene
scintillator tiles, as shown in \cref{fig:Tile}. The scintillator tiles
are wrapped in reflective foil by a robotic procedure prior to automatic placement
on the HBU board with assembled photon sensors.
The full hadronic prototype will see
first beam in 2018.

\section{Conclusion}
The engineering prototype of the AHCAL is an important tool to verify the technologies
developed in the CALICE collaboration for the construction of a highly granular
hadronic calorimeter for a future linear collider detector.
As one step towards this goal, a test-beam
campaign was carried out in order to test the performance of the prototype in a
strong magnetic field. Data of muons and electrons at different energies were
taken with and without a magnetic field of \SI{1.5}{\tesla}.

A broadening of the
shower due to the magnetic field is clearly visible. The detector response with the
magnetic field is increased by roughly \num{10}\%. This effect is a combination
of an increased light yield of the scintillator, more observed hits in the event
and a longer path of the particles traversing the scintillator due to the bending
in the magnetic field. The increase in light yield of the scintillators of roughly
\num{4}\% is in agreement with previous observation in literature. No significant
influence of the magnetic field on the time measurement is being observed.
As a next step the construction of a large prototype with \num{23000} channels is currently
ongoing and will be tested in 2018.






%

\bibliographystyle{IEEEtran.bst}
\bibliography{CALICE}

\begin{thebibliography}{1}
\providecommand{\url}[1]{#1}
\csname url@samestyle\endcsname
\providecommand{\newblock}{\relax}
\providecommand{\bibinfo}[2]{#2}
\providecommand{\BIBentrySTDinterwordspacing}{\spaceskip=0pt\relax}
\providecommand{\BIBentryALTinterwordstretchfactor}{4}
\providecommand{\BIBentryALTinterwordspacing}{\spaceskip=\fontdimen2\font plus
\BIBentryALTinterwordstretchfactor\fontdimen3\font minus
  \fontdimen4\font\relax}
\providecommand{\BIBforeignlanguage}[2]{{%
\expandafter\ifx\csname l@#1\endcsname\relax
\typeout{** WARNING: IEEEtran.bst: No hyphenation pattern has been}%
\typeout{** loaded for the language `#1'. Using the pattern for}%
\typeout{** the default language instead.}%
\else
\language=\csname l@#1\endcsname
\fi
#2}}
\providecommand{\BIBdecl}{\relax}
\BIBdecl

\bibitem{Thomson:2009rp}
M.~Thomson, ``{Particle Flow Calorimetry and the PandoraPFA Algorithm},''
  \emph{Nucl.Instrum.Meth.}, vol. A611, pp. 25--40, 2009.

\bibitem{Adloff:2010hb}
C.~Adloff \emph{et~al.}, ``{Construction and Commissioning of the CALICE Analog
  Hadron Calorimeter Prototype},'' \emph{JINST}, vol.~5, p. P05004, 2010.

\bibitem{Adloff:2012gv}
------, ``{Hadronic energy resolution of a highly granular scintillator-steel
  hadron calorimeter using software compensation techniques},'' \emph{JINST},
  vol.~7, p. P09017, 2012.

\bibitem{Adloff:2011ha}
------, ``{Tests of a particle flow algorithm with CALICE test beam data},''
  \emph{JINST}, vol.~6, p. P07005, 2011.

\bibitem{Simon:2010hf}
F.~Simon and C.~Soldner, ``{Uniformity Studies of Scintillator Tiles directly
  coupled to SiPMs for Imaging Calorimetry},'' \emph{Nucl.Instrum.Meth.}, vol.
  A620, pp. 196--201, 2010.

\bibitem{Liu:2015cpe}
\BIBentryALTinterwordspacing
Y.~Liu \emph{et~al.}, ``{A Design of Scintillator Tiles Read Out by
  Surface-Mounted SiPMs for a Future Hadron Calorimeter},'' in \emph{{IEEE NSS
  2014}}, 2015. [Online]. Available:
  \url{http://inspirehep.net/record/1410633/files/arXiv:1512.05900.pdf}
\BIBentrySTDinterwordspacing

\bibitem{di2013spiroc}
D.~Lorenzo \emph{et~al.}, ``Spiroc: design and performances of a dedicated very
  front-end electronics for an ilc analog hadronic calorimeter (ahcal)
  prototype with sipm read-out,'' \emph{JINST}, vol.~8, no.~01, p. C01027,
  2013.

\bibitem{bertolucci1987MagneticField}
Bertolucci \emph{et~al.}, ``Influence of magnetic fields on the response of
  acrylic scintillators,'' \emph{Nucl. Instr. Meth. Phys. Res. A}, vol. 254,
  no.~3, pp. 561--562, 1987.

\bibitem{bertoldi1997scintillators}
Bertoldi \emph{et~al.}, ``Scintillators in magnetic fields up to 20 t,''
  \emph{Nucl. Instr. Meth. Phys. Res. A}, vol. 386, no. 2-3, pp. 301--306,
  1997.

\end{thebibliography}




\end{document}